\documentclass[twocolumn,%
  aps,superscriptaddress,%
  amsmath,amssymb,amsfonts,floatfix,
  eqsecnum,prx,10pt]{revtex4-2}

\usepackage[utf8]{inputenc}
\usepackage[T1]{fontenc}
\usepackage{lmodern}
\usepackage{microtype}

\usepackage[toc,page]{appendix}
\usepackage{graphicx} % Required for inserting images

\usepackage{mathtools}
\usepackage{amsthm}
\usepackage{braket}
\usepackage{tikz-cd}
\usepackage{cancel}
\usepackage{bbold}
\usepackage{hyperref}
\hypersetup{
    colorlinks=true, 
    allcolors=blue,
}
\usepackage[capitalise]{cleveref}
\usepackage{hhline}
\usepackage{mleftright}

\usepackage{mycommands}

\begin{document}

\begin{abstract}
    Conventional theories for Mott insulators involve well-localized electronic orbitals. This picture fails in the presence of topological obstructions in Chern bands which prevent the formation of exponentially localized orbitals and are instead often viewed by analogy to Landau levels. Here, we show that strong interactions in fractionally-filled topological bands realize a class of Mott insulating states on emergent topological Kondo lattices. These can be naturally described by a partial Wannier basis of $N-1$ exponentially-localized orbitals for an $N$-band manifold with net non-zero Chern number. Choosing a gauge which breaks translation symmetry, we construct a supercell basis which segregates an isolated topological band into well-localized supercell Wannier orbitals that can host the local moments of a Mott insulating state, as well as an itinerant power-law-localized orbital that retains the topological obstruction. Together, they constitute a spontaneous Kondo lattice. We study fractionally-filled interacting quantum spin Hall bands in the Bernevig-Hughes-Zhang model as well as twisted bilayers of MoTe\textsubscript{2} at $\nu = -1/3$ filling. We demonstrate that strong short-ranged Coulomb interactions can stabilize a new class of topological Mott insulating states with broken translation symmetry and antiferromagnetic order, which compete with fractional Chern insulating states. Our results predict a new scenario for time-reversal-symmetric interacting topological bands in solids, beyond conventional Landau level paradigms for the fractional quantum Hall effect.
\end{abstract}

\title{Supercell Wannier Functions and Emergent Kondo Lattices in Topological Bands}
\author{Brandon Monsen}
\email{bmonsen@sas.upenn.edu}
\affiliation{Department of Physics and Astronomy, University of Pennsylvania, Philadelphia, PA 19104, USA}
\author{Martin Claassen}
\email{claassen@sas.upenn.edu}
\affiliation{Department of Physics and Astronomy, University of Pennsylvania, Philadelphia, PA 19104, USA}
\date{\today}
\maketitle

%-----------------------------------------------------------------------------------------
%
% INTRODUCTION
% 
%-----------------------------------------------------------------------------------------

\section{Introduction}

The recent discovery of flat topological bands in moir\'e heterostructures holds tremendous promise for realizing exotic new states of matter with useful properties. Thus far, spurred by analogies to fractionally-filled Landau levels~\cite{regnault_fractional_2011,sun_nearly_2011,wu_zoology_2012,bernevig_emergent_2012,mcgreevy_wave_2012,wu_bloch_2013,Estienne2023a}, 
groundbreaking experimental and theoretical efforts have primarily focused on the realization of fractional quantum Hall (FQH) states in moir\'e bands without external magnetic fields, provided that time-reversal symmetry is spontaneously broken. Fractional Chern insulators (FCIs) have been observed in twisted bilayers of MoTe$_2$~\cite{cai_signatures_2023, redekop_direct_2024, park_observation_2023, zeng_thermodynamic_2023} and rhombohedral pentalayer graphene~\cite{Lu2024b}, raising exciting new prospects for engineering non-Abelian states, quantum Hall crystalline phases, and beyond~\cite{anderson_programming_2023}.

However, in stark contrast to the conventional quantum Hall paradigm, moir\'e heterostructures such as twisted bilayers of MoTe\textsubscript{2} are \textit{a priori} characterized by \textit{time-reversed} copies of quantum spin Hall (QSH) bands~\cite{wu_topological_2019,devakul_magic_2021}, and moreover have tunable quantum-geometric properties that can differ profoundly from both ideal Landau levels and conventional Mott-Hubbard settings ~\cite{roy2014,claassen2015,jackson2015,Lee2017,wang2021,ledwith2023,liu2024}. This raises intriguing questions regarding new states of matter from an interplay of topology, spontaneous symmetry breaking, and strong electronic interactions beyond the Landau-level paradigm. An example is the quantum anomalous Hall crystal, which spontaneously breaks both time reversal and translation symmetry and results in a topological state with Chern number $\Chern = 1$ despite being at fractional filling~\cite{sheng_quantum_2024, perea-causin_quantum_2024, zhou_new_2024}. 

Canonical prescriptions for the quantum Hall problem cease to apply if the moir\'e band displays strong violations of ideality conditions for the quantum metric~\cite{roy2014,claassen2015,Lee2017,wang_exact_2021} or a deviation from vortexability~\cite{ledwith2023}. On the other hand, a topological obstruction in Chern bands can prevent the formation of conventional Mott insulators. Here, the root cause is an inability to form exponentially localized Wannier functions \cite{PhysRevLett.98.046402,PhysRevB.74.235111} unless time-reversal symmetry or other protecting symmetries are implemented non-locally \cite{soluyanov_wannier_2011}, which in turn precludes a local Hubbard-like description of the partially filled band. For instance, a fragile topological obstruction protected by $\mathcal{C}\textsubscript{2}\mathcal{T}$ in twisted bilayer graphene can be circumvented by extended Wannier orbitals with a fidget-spinner shape, which however cease to remain localized within the unit cell~\cite{Po2018a,koshino_maximally_2018,carr_derivation_2019}. Consequently, correlated insulating states in topological bands are not easily viewed in terms of Mott-Hubbard models and localized charges even for strong and well-screened Coulomb interactions, provided that other bands remain energetically separated. 

In this work, we introduce a new class of Mott insulating states in fractionally filled interacting topological bands that can be explained by a \textit{partial} Wannier basis. The partial Wannier basis provides a well-localized but incomplete set of orbitals for Mott insulators in topological bands by choosing a gauge that violates translation symmetry. As an immediate consequence, we demonstrate that strong interactions in fractionally filled topological bands can realize an \textit{emergent} topological Kondo lattice, composed of a spontaneously generated regular arrangement of localized magnetic moments (Wannier functions) in the presence of itinerant extended electronic states that retain the topological character of the host band. The resulting topological Mott-insulating states of matter spontaneously break translation symmetry, possibly exhibit magnetic order, but can intrinsically remain spin-unpolarized, in contrast to quantum anomalous Hall crystals in Landau levels and Chern bands. They appear as a competing phase to time-reversal-breaking FCIs and become favored if the band supports a tightly-localized partial Wannier basis.

We illustrate the formation of emergent Kondo lattices and Mott insulators for interacting fractionally-filled spin Chern bands in the Bernevig-Hughes-Zhang (BHZ) model~\cite{Bernevig2006a} as well as in twisted bilayers of MoTe\textsubscript{2} \cite{wu_topological_2019,cai_signatures_2023}. Our results provide a new scenario for understanding and predicting topological correlated states without time-reversal-symmetry breaking in twisted transition-metal dichalcogenide heterostructures and beyond.

%-----------------------------------------------------------------------------------------
%
% WANNIER CONSTRUCTION
% 
%-----------------------------------------------------------------------------------------

\begin{figure*}[hbt!]
    \centering
    \includegraphics[width = \linewidth]{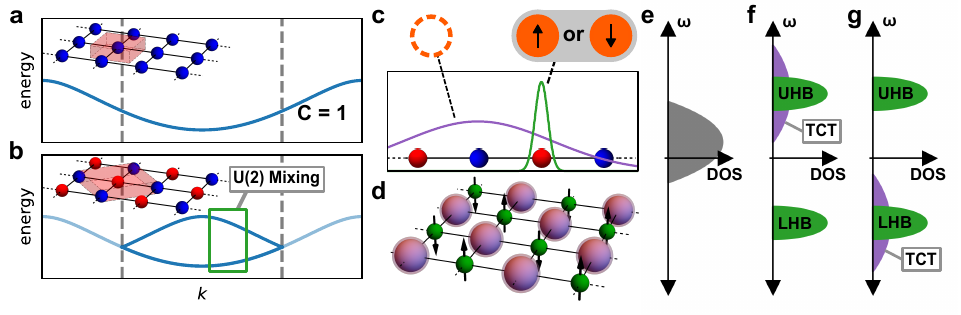}
    \caption{\captiontitle{Supercell Wannier functions for topological bands.} \textbf{(a)} An isolated Chern band has a topological obstruction, preventing the construction of exponentially-localized Wannier functions [inset: example square lattice unit cell]. %An isolated Chern band on a lattice model. The unit cell is shown in the red box.
    \textbf{(b)} Redefining the model on an enlarged unit cell [inset] shrinks the Brillouin zone and folds the Chern band. Red and blue lattice sites are no longer identified. A supercell Wannier construction now involves two bands. %A $U(2)$ rotation permits the construction of one exponentially-localized Wannier function, as well as one power-law-localized orbital that retains the topological character of the original band.
    \textbf{(c)} $U(2)$ mixing permits the construction of one exponentially-localized supercell Wannier orbital (SCWO, green), as well as one topological power-law orbital (TPLO, purple) that retains the topological character of the original band. Strong short-ranged Coulomb repulsion naturally favors a Mott-like state with a single spin $\uparrow$ or $\downarrow$ electron localized per SCWO, as well as empty or dilutely-filled TPLOs.
    \textbf{(d)} Local magnetic moments in tightly-localized SCWOs (green) are coupled to the itinerant band composed of TPLOs (purple), forming an emergent topological Kondo lattice.
    \textbf{(e)} Schematic spectral function before interactions are turned on, depicting a metallic state at fractional filling.
    \textbf{(f)} Schematic spectral function deep in the topologically-trivial Mott phase at $\nu = 1/2$ filling. 
    \textbf{(g)} Schematic spectral function deep in the topological Mott phase with a filled topological charge-transfer (TCT) band at $\nu = 3/2$.
    }
    \label{fig:wannier-heuristic}
\end{figure*}

\section{Partial Wannier Basis and Emergent Kondo Lattices}
\label{sec:supercell-wannier-function}

Wannier orbitals form the single-particle basis for tight-binding descriptions of interacting electrons in partially filled valence bands. A Wannier basis for a single band is a complete orthogonal set of exponentially localized orbitals $\phi_{\R}(\r)$ that are related via lattice translations $\phi_{\R}(\r) = \phi_{0}(\r-\R)$. They can be defined as Fourier transforms $\phi_{\R}(\r) = \frac{1}{\sqrt{N}} \sum_{\k} e^{-i\k(\R-\r)} u_\k(\r)$ where $u_\k(\r)$ denotes the cell-periodic part of the Bloch state and $N$ is the number of unit cells. Wannier functions are not unique~\cite{marzari_maximally_1997}. For a single band, fixing an intrinsic $U(1)$ gauge freedom $u_\k(\r) \to e^{i \chi_\k} u_\k(\r)$ determines the shape of the orbital. Exponential localization requires a smooth and periodic gauge choice across the entire Brillouin zone, which is impossible in a band with non-zero Chern number~\cite{marzari_maximally_1997,brouder_exponential_2007}. For $N>1$ bands, a Wannier basis with $N$ exponentially localized orbitals can be recovered by enlarging the gauge group $U(1) \rightarrow U(N)$, provided that the net Chern number is zero~\cite{winkler_smooth_2016}.

\subsection{Heuristic}

The starting point for this work is the observation that a \textit{partial} Wannier basis with $N-1$ exponentially localized orbitals is possible even for topological $N$-band manifolds with \textit{non-zero} net Chern numbers $\Chern$. Heuristically, suppose that these bands are spanned by a basis of $n=1,\dots,N$ Bloch states $u_{n\k}(\r)$ which are periodic and smooth everywhere except for (gauge-dependent) isolated points in the Brillouin zone (BZ). Now, a $U(N)$ rotation $w_{m\k}(\r) = \sum_n U_{\k}^{nm} u_{n\k}(\r)$ can selectively shift all phase vortex nonanalyticities to a \textit{single state} ($m=N$), while making the remaining $N-1$ states ($m \leq N-1$) smooth over the entire BZ. Upon Fourier transformation $W_{m\R}(\r) = \frac{1}{\sqrt{N}} \sum_{\k} e^{-i\k(\R-\r)} w_{m\k}(\r)$, these states describe $N-1$ exponentially localized Wannier orbitals, as well as a single non-localizable topologically-obstructed orbital that encodes the Chern number $\Chern$ of the original band manifold, and is characterized by power-law decay as a function of $\r$.

In fractionally filled topological bands, a partial Wannier basis is a natural basis for understanding translation-symmetry-breaking Mott or charge-ordered phases. To see this, we first ignore spin and consider an isolated Chern band at half filling [Fig.~\ref{fig:wannier-heuristic}(a)]. Imposing an artificially doubled unit cell leads to two folded bands that touch at a reduced Brillouin zone (rBZ) boundary [Fig.~\ref{fig:wannier-heuristic}(b)]. The net Chern number remains unchanged. However, the folded band manifold now comprises two supercell Bloch states with an enlarged gauge group $U(2)$ from mixing between the folded bands. This can be exploited to construct a partial Wannier basis for the supercell, composed of one exponentially localized Supercell Wannier orbital (SCWO) and a second topological power-law-localized orbital (TPLO) that retains the topological obstruction of the original Chern band. The construction can be thought of as a gauge choice that does not respect all lattice translations, in analogy to time-reversal-non-respecting gauges for $\mathbb{Z}_2$ topological insulators~\cite{soluyanov_wannier_2011}. The resulting basis does not explicitly break translation symmetry, which remains encoded in a nonlocal transformation between the supercell Wannier orbital and the second power-law-localized orbital.

The effect of strong Coulomb repulsion between electrons in topological bands becomes particularly transparent in this basis. A screened local interaction will necessarily remain short-ranged for pairs of electrons in SCWOs. In contrast, two electrons in spatially-separated TPLOs will incur a sizable interaction energy penalty even for well-screened Coulomb interactions via overlapping power-law tails. In a half-filled flat Chern band, the Coulomb interaction energy is therefore readily minimized by placing a single electron per SCWO while leaving the TPLOs unoccupied [Fig. \ref{fig:wannier-heuristic}(c)]. The resulting charge-ordered state spontaneously breaks translation symmetry and has a charge gap, with gapped density excitations from moving an electron from an SCWO into a TPLO, or into a doubly-occupied SCWO.

Including spin leads to an entirely new class of Mott states. In time-reversal-symmetric spin Chern bands, the partial supercell Wannier construction results in a pair of identical real-space orbitals for spin $\uparrow$ and $\downarrow$. If the band in Fig. \ref{fig:wannier-heuristic}(a) is a Kramers pair of QSH bands, then the same heuristic would obtain one orbital in Fig. \ref{fig:wannier-heuristic}(c) per spin. At filling $\nu = 1/2$ ($1/4$ filling per spin per unit cell), Coulomb interaction energies are still minimized by placing a single electron in the SCWO, while the TPLOs remain unoccupied. Such a Mott state appears as a translation-symmetry-breaking phase below a finite temperature $T_c$ which is proportional to effective Hubbard interactions in the SCWOs. These interactions can be computed by rotating and projecting the screened Coulomb interaction vertex into the supercell Wannier basis. Importantly, the electrons in the SCWOs now form local magnetic moments. These can order magnetically at much lower temperatures $T_m \ll T_c$, governed by the effective spin exchange interactions between SCWOs. However, more exotic quantum spin liquid phases are possible~\cite{savary2016quantum,broholm2020quantum,RevModPhys.89.025003}.

The single-particle spectrum of the resulting Mott insulating state at $\nu = 1/2$ is is characterized by lower and upper Hubbard bands forming from the SCWOs, as well as an empty topological charge transfer (TCT) band of TPLOs, depicted in Fig. \ref{fig:wannier-heuristic}(f). Adding a few extra electrons leads to a dilute filling of the TPLOs, which take the role of itinerant topological charge-transfer or metallic states. The resulting interacting problem therefore spontaneously generates an emergent Kondo lattice [Fig. \ref{fig:wannier-heuristic}(d)], composed of localized magnetic moments coupled to an empty or dilutely-filled itinerant topological band of electrons or holes.

Topologically non-trivial Mott insulating states can arise as the TPLOs become occupied. For example, the Mott insulating state at $\nu = 3/2$ filling of the QSH bands [Fig. \ref{fig:wannier-heuristic}] can be viewed as a ``$\nu = 1/2$ hole'' Mott insulator. From the perspective of electrons, the TCT band of TPLOs is now fully occupied. The resulting Mott state is an integer quantum spin Hall insulator that also spontaneously breaks translation symmetry [Fig. \ref{fig:wannier-heuristic}(g)] -- we dub this state a ``quantum spin Hall crystal'', in analogy to quantum anomalous Hall crystals in Chern bands~\cite{tesanovic_hall_1989, murthy_hall_2000, sheng_quantum_2024, perea-causin_quantum_2024, zhou_new_2024}. Notably, if the quantum spin Hall crystal is topologically protected solely by time-reversal symmetry, it exists only at temperatures $T_m < T < T_c$ above a magnetic ordering temperature.

This heuristic also holds at other filling fractions. Given a fractional commensurate filling, the general principle is to choose a supercell that supports the expected charge ordering pattern, construct a partial supercell Wannier basis, and study the propensity for the formation of an emergent Kondo lattice as a function of the localization properties of the resulting SCWOs. At $\nu = 1/3$ filling on a triangular lattice, a $\sqrt{3} \times \sqrt{3}$ charge ordering can be described by a tripled unit cell with two SCWOs and one TPLO, where the SCWOs need not be related by simple translations. Importantly, multiple supercell constructions are possible for the same filling. For example, at $\nu = 1/2$ in a highly anisotropic rectangular-lattice model, it is also possible to have a stripe ordering pattern described by SCWOs in a doubled $2 \times 1$ unit cell that breaks rotational symmetry. The supercell Wannier construction therefore provides a powerful heuristic for the stability of competing states and emergent Kondo lattices in fractionally-filled topological bands, and for characterizing their topology.

\subsection{Supercell Wannier Functions}

In practice, a natural way to construct a smooth gauge and a partial Wannier basis for an artificially enlarged unit cell is through the perspective of a supercell Hamiltonian. Suppose the original crystal or moir\'e lattice has primitive vectors $\a_{1}$, $\a_{2}$, and reciprocal lattice vectors $\b_{1}$, $\b_{2}$. A supercell is defined by new translation vectors $\A_{1} = m_{11} \a_{1} + m_{12} \a_{2}$ and $\A_{2} = m_{21} \a_{1} + m_{22} \a_{2}$ where $m_{11}, m_{12}, m_{21}, m_{22} \in \mathbb{Z}$. For moir\'e heterostructures, this describes a supercell of moir\'e supercells. The `broken' lattice is then $L_{b} = \{n_{1} \A_{1} + n_{2} \A_{2} ~:~ n_{1}, n_{2} \in \mathbb{Z}\}$. Each such supercell contains $N \equiv \det\{ m_{ij} \}$ original unit cells. The locations of the original unit cells within the supercell define a set $\mathcal{T}$ of $l=1,\dots,N$ tiling vectors $\t_l = n_{l,1} \a_1 + n_{l,2} \a_2$ with $n_{l,1}, n_{l,2} \in \mathbb{Z}$, including the zero vector. The sets $L(\t) := \{\R + \t : \R \in L_{b} \}$ therefore form a partition of the original lattice. In momentum space, the corresponding reduced Brillouin zone (rBZ) for the supercell has reciprocal lattice vectors $\mathbf{B}_1 = (m_{22} \b_{1}  - m_{21} \b_{2}) / N$ and $\mathbf{B}_2 = (-m_{12} \b_{1} + m_{11} \b_{2}) / N$. The rBZ similarly tiles the original Brillouin zone (BZ), which defines a set $\mathcal{G}$ containing $N$ reciprocal tiling vectors $\g_l$. The real-space and reciprocal tiling vectors satisfy $(1/N) \sum_{\t} e^{i \t \cdot \g} = \delta_{\g,\mathbf{0}}$. We denote momenta in the original BZ and rBZ as $\k$ and $\K$, respectively.

We now construct a supercell Bloch Hamiltonian $\mathbf{H}(\K)$ that makes the real-space tiling explicit. The transformation between original and supercell degrees of freedom amounts to a shift from original lattice vectors to a pair consisting of a broken lattice vector and a tiling vector. Their correspondence comes from the previously mentioned lattice partition. Explicitly, we have 
\begin{align}
    \cD_{\R + \t, \alpha} = \CD_{\R, \t, \alpha},
\end{align}
where $\alpha$ is an enumeration of internal degrees of freedom such as orbital and original sublattice. This transformation amounts to moving lattice sites that correspond to tiling vectors in a supercell into a new effective sublattice degree of freedom. From here, transforming to supercell degrees of freedom is a matter of direct substitution.

Consider a Bloch Hamiltonian that is defined on the original lattice. Its matrix representation $\mathbf{h}(\k)$ is an $M \times M$ tight-binding Bloch Hamiltonian for the original unit cell, describing $M$ orbitals and sublattices. Now, every momentum in the original BZ $\k = \K + \g$ is covered by an rBZ momentum $\K$ and a reciprocal tiling vector $\g$. Fourier transforming the real-space operators on their respective unit cells gives a relationship between the fermionic creation operators in the original ($\cD_{\k,\alpha}$) and reduced ($\CD_{\K,\t,\alpha}$) BZ:
\begin{align}
    \label{eqn:supercell-creation-operator-momentum-space}
    \cD_{\K + \g,\alpha} &= \frac{1}{\sqrt{|\mathcal{T}|}} \sum_{\t \in \mathcal{T}} e^{i \g \cdot (\t + \d_{\alpha})} \CD_{\K, \t, \alpha}
\end{align}
A tight-binding model $\hat{H} = \sum_{\k,\alpha, \beta} h_{\alpha\beta}(k)~ \cD_{\k, \alpha} \c_{\k,\beta}$ then has a corresponding supercell model $\hat{H} = \sum_{\k,\alpha,\beta} \sum_{\t_{1},\t_{2} \in \mathcal{T}} H_{\t_{1}, \alpha, \t_{2}, \beta}(\K) \CD_{\K, \t_{1}, \alpha} \C_{\K, \t_{2}, \beta}$.
As a matrix, $\H(\K)$ satisfies
\begin{align}
    \label{eqn:supercell-bloch-hamiltonian}
    \H(\K) = \sum_{\g \in \mathcal{G}} \T\adj(\g) \left[h(\K + \g)_{\alpha, \alpha'} \otimes \mathbf{O}_{\t,\t'} \right] \T(\g),
\end{align}
where $\mathbf{O}_{\t,\t'}$ is a matrix of ones, acting on each original unit cell equally $O_{\t, \t'} = 1$ and 
\begin{align}
    \label{eqn:supercell-momentum-embedding-operator}
    T(\g)_{\alpha \t, \alpha' \t'} = \delta_{\alpha \alpha'} \delta_{\t \t'} e^{-i \g \cdot (\t + \d_{\alpha})}
\end{align}
is the tiled supercell embedding operator. The Bloch states are then inherited from the parent Hamiltonian as
\begin{align}
    \label{eqn:supercell-bloch-states}
    \ket{U^{\g}(\K)} = \T\adj(\g) \bigoplus_{\t \in \mathcal{T}} \ket{u(\K + \g)},
\end{align}
where $\ket{u(\k)}$ is the Bloch state for the original topological band.
Physically, these states describe the unbroken Bloch states duplicated for each tiling vector location and properly embedded by $\T$. %These calculations are performed in more detail in the supplement.

To construct the SCWOs for a partial Wannier basis, a smooth gauge choice must now be implemented. However, the supercell Bloch states $\ket{U^{\g}(\K)}$ are discontinuous at the rBZ boundary. Crossing the rBZ boundary places $\K$ in a different tiling vector's partition region of the full Brillouin zone, so a smooth interpolation will change the $\g$ index of the supercell Bloch state. A simple strategy to construct a smooth gauge is to instead consider projectors~\cite{marzari_maximally_1997} onto the set of supercell bands that form from folding a topological band into the rBZ:
\begin{align}
    \label{eqn:folded-projector}
    \hat{P}(\K) = \sum_{\g} \ket{U^{\g}(\K)} \bra{U^{\g}(\K)}.
\end{align}
Physically, this object projects into the isolated Chern band but is expressed in supercell degrees of freedom. It is smooth and periodic in the rBZ, provided that the original topological band is spectrally isolated.

A partial Wannier basis for a supercell containing $N$ original unit cells can now be constructed by choosing a set of $m=1,\dots,N-1$ smooth trial wavefunctions $\ket{\psi_m(\K)}$ such that the image $\hat{P}(\K) \ket{\psi_m(\K)}$ is smooth and periodic in the supercell but may violate the translation symmetry of the original lattice. The projected trial wave functions $\hat{P}(\K) \ket{\psi_m(\K)}$ must now be orthonormalized, while ensuring that they remain smooth in the rBZ. A simple recipe is to use the iterative Gram-Schmidt process. Suppose that the projected first trial wave function $\ket{\psi_1}$ does not vanish anywhere in the rBZ such that
\begin{align}
    \ket{\Psi_1(\K)} = \frac{ \hat{P}(\K) \ket{\psi_1(\K)} }{ \bra{\psi_1(\K)} \hat{P}(\K) \ket{\psi_1(\K)} }
\end{align}
remains globally smooth. This state defines an SCWO after Fourier transform. For $N>2$, an $m^{\rm th}$ orthonormal state ($m < N$) can be defined iteratively by choosing another trial wave function, projecting, and orthonormalizing with respect to $\ket{\Psi_{m'}(\K)}$ for $m' < m$. For suitable choices of $\ket{\psi_m(\K)}$, this defines $N-1$ smooth projected orthonormal trial states. The corresponding real-space SCWOs in supercell $\R_0$ are now given by the Fourier transform
\begin{align}
    \ket{W_{m,\R_0}(\R)} = \frac{1}{\sqrt{L}} \sum_{\K} e^{i\K(\R - \R_0)} \ket{\Psi_m(\K)}
\end{align}
which decays exponentially with distance $\R - \R_0$. These SCWOs form an orthonormal (but incomplete) Wannier basis for the topological band. Maximally-localized Wannier orbitals with a minimized mean squared real-space spread can now be computed in a standard way \cite{PhysRevB.56.12847}, via further optimization of the trial wave functions or unitary rotations among the states $\ket{\Psi_m(\K)}$.

However, the topological obstruction of the original Chern band prevents the $N^{\rm th}$ orbital from being exponentially localized for any choice of trial wave function. In the Gram-Schmidt process, the final $N^{\rm th}$ projected trial state must vanish at discrete points in the rBZ after orthogonalization, and hence acquires phase vortex-like nonanalyticities after normalization. After Fourier transform, the state $\ket{W_{N, \R_0}}$ therefore defines the TPLO and will decay as a power-law with distance $\R - \R_0$.

\section{Emergent Kondo Lattices in Fractionally-Filled Topological Bands}

We now verify the intuition laid out in the previous section by studying fractionally-filled quantum spin Hall bands using exact diagonalization (ED). We explicitly demonstrate the emergence of topological Kondo lattices in isolated, interacting topological bands, and show that the SCWOs/TPLO form a natural basis for describing antiferromagnetic and ferromagnetic Mott insulating ground states. This demonstration is done for two scenarios. The first is a interacting Bernevig-Hughes-Zhang (BHZ) model~\cite{Bernevig2006a} on a square lattice at $1/2$ filling ($1/4$ electrons per spin per unit cell) and $3/2$ filling. The second candidate scenario is given by twisted bilayers of MoTe\textsubscript{2}. Here, we show that emergent topological Kondo lattices with antiferromagnetic order form spontaneously at $-1/3$ filling and twist angles $\theta \gtrsim 4^\circ$, realizing a new topological phase that is a competing state to FCIs.

Theoretically, we confirm the formation of spontaneous Kondo lattices in both cases by finding parameters where a Mott insulating state with spontaneously broken translation symmetry forms. These initial calculations will know nothing about the partial Wannier basis. Next, orbitals for the spontaneous Kondo lattice are constructed via the projective supercell construction [Eq. (\ref{eqn:folded-projector})]. The supercell is chosen based on the charge ordering found in the ED calculations, though in principle one could infer the correct supercell from filling fraction and symmetries. The final step is to determine how well the low energy many-body eigenstates are captured via local magnetic moments in the supercell Wannier basis.

\subsection{BHZ Model}

\begin{figure*}[tb!]
    \centering
    \includegraphics[width = \linewidth]{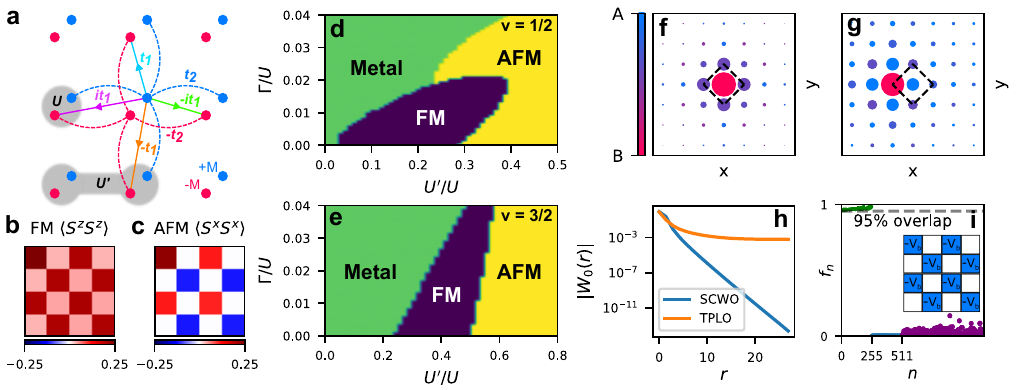}
    \caption{\captiontitle{Emergent Kondo Lattices in the interacting BHZ Model.} \textbf{(a)} Schematic of the hopping amplitudes $t_1$, $t_2$, interaction terms $U$ and $U'$, and mass $M$. Blue and red dots represent an $s-$ and $p-$like orbital per site. \textbf{(b)} Representative $S^{z}S^{z}$ correlation function from the FM phase which is identical for both $\nu = 1/2$ and $\nu = 3/2$ ($1/2$ hole filled). \textbf{(c)} Analogous representative $S^{x}S^{x}$ correlation function for the AFM region. \textbf{(d)} The phase diagram at $\nu = 1/2$ filling ($1/4$ electrons per spin per unit cell), as a function of the artificially tuned valence band width $\Gamma$ and nearest-neighbor interaction $U'$. \textbf{(e)} Analogous phase diagram for $\nu = 3/2$. \textbf{(f)}, \textbf{(g}) Plots of the SCWO, TPLO orbitals, showing orbital character (color of dots) and amplitude (area of dots) on each site. The supercell is outlined in black dashed lines. \textbf{(h)} The amplitude of the SCWO/TPLO as a function of distance along the $A_{1}$ direction. \textbf{(i)} The projection $f_{n}$ for the first $n = 1200$ many-body eigenstates. A weak bias field (inset) is applied to single out one translation-symmetry-breaking pattern.}
    \label{fig:BHZ}
\end{figure*}

We first consider the square-lattice BHZ model with a fractionally-filled valence band. On the square lattice, the partial Wannier basis construction suggests that $1/2$ or $3/2$ valence band filling ($1/4$ or $3/4$ electrons per spin per unit cell) can stabilize the densest emergent Kondo lattice with a doubled unit cell and preserved $C_4$ rotation symmetry.

The interacting Hamiltonian is shown schematically in figure \cref{fig:BHZ}a. It can be written as
\begin{align}
    \hat{H} = \sum_{\k} \hat{\boldsymbol{c}}_{\k}^\dag \cdot \begin{bmatrix}
        \boldsymbol{h}(\k) & 0 \\
        0 & \boldsymbol{h}^{*}(-\k)
    \end{bmatrix} \cdot \hat{\boldsymbol{c}}_{\k} + \frac{1}{2} \sum_{\q} V(\q) \hat{\rho}_\q \hat{\rho}_{-\q}  \label{eq:BHZfull}
\end{align}
where $\hat{\boldsymbol{c}}_{\k} = [ \c_{\k s \uparrow}, \c_{\k p \uparrow}, \c_{\k s \downarrow}, \c_{\k p \downarrow} ]^\top$ are fermion operators for Bloch electrons in $s$- and $p$-like orbitals per site, $\boldsymbol{h}(\k)$ is the single-particle Bloch Hamiltonian, and $V(\q)$ parametrizes Coulomb interactions where $\hat{\rho}_\q = (1 / \sqrt{L}) \sum_{\k} \sum_{\sigma} \sum_{\alpha=s,p} \cD_{\k+\q,\alpha\sigma} \c_{\k,\alpha\sigma}$ is the density operator and $L$ is the lattice size. The Bloch Hamiltonian
\begin{equation}
    \begin{aligned}
        \boldsymbol{h}(k) = &-2t_{1} \left( \sin (k_{x}) \tau_{x} + \sin(k_{y}) \tau_{y} \right) \\
        &+ \left[M - 2t_{2} (\cos k_{x} + \cos k_{y}) \right] \tau_{z},
    \end{aligned}
\end{equation}
describes two spin-degenerate bands, which have a non-zero spin Chern number for the choice of parameters $t_{1}/M = t_{2}/M = 0.5$~\cite{Bernevig2006a}. The Pauli matrices $\tau_{x}, \tau_{y}, \tau_{z}$ act on orbital degrees of freedom. We choose density-density interactions $V(\q) = U + 2 U' \left[ \cos(q_{x}) + \cos(q_{y}) \right]$ with an on-site $U$ and nearest-neighbor contribution $U'$ that act equally on both orbitals and spins. For Coulomb interactions smaller than the single-particle band gap, Eq. (\ref{eq:BHZfull}) can be projected into the valence band in analogy to Landau level projections. The ratio of bandwidth $\Gamma$ and interactions $U$ of the lower band after projection can also be artificially tuned, which can be thought of as adding progressively longer-ranged hoppings to affect the dispersion while keeping the Bloch states and Chern number identical. 

Fig. \ref{fig:BHZ}(d) and (e) show the phase diagrams for $1/2$ and $3/2$ filling of the quantum spin Hall valence band, obtained using ED on a $4 \times 4$ lattice with periodic boundary conditions. The phase boundaries are identified via second derivatives of the ground state energy with model parameters, ground state degeneracies, spin polarization, and correlation functions. At $1/2$ filling in the flat-band limit ($\Gamma = 0$), a metallic region immediately gives way to a ferromagnetic-ordered Mott insulator for finite nearest-neighbor Coulomb repulsion $U'/U$. 
The FM region shows clear signatures of spontaneous translation symmetry breaking via their checkerboard pattern in both spin [Fig. \ref{fig:BHZ}(b)] and charge correlation functions. This is consistent with the formation of an emergent Kondo lattice, composed of magnetic moments from charges that are localized on a checkerboard of SCWOs, and an unoccupied itinerant band of TPLOs, as expected from partial Wannier basis heuristics.

Intriguingly, a transition to an antiferromagnetic Mott state without net spin polarization appears as $U'/U$ or bandwidth $\Gamma / U$ is increased. This can be understood to arise from a competition between direct (Coulomb-interaction-mediated) exchange and superexchange interactions between the magnetic moments, tunable by bandwidth, the range of Coulomb interactions, and the localization properties of the SCWOs.

The inequivalency of $\nu = 1/2$ and $\nu = 3/2$ is a direct result of explicitly broken particle-hole symmetry~\cite{lauchli_hierarchy_2013} in the spin Chern bands. A particle hole transformation of the projected interaction generates an effective dispersion for holes. Consequently, a metallic phase persists up to a finite interaction strength even for a vanishing bare single-particle bandwidth $\Gamma/U = 0$. ED results are again consistent with the heuristic picture [Fig. \ref{fig:wannier-heuristic}] of placing one electron in each SCWO for a checkerboard supercell. 

A key topological distinction between Mott insulators at $1/2$ and $3/2$ filling is the occupancy of the TPLOs. For $\nu = 1/2$, all TPLOs are unoccupied and the Mott insulator remains topologically trivial. In contrast, for $\nu = 3/2$, the ``topological charge transfer'' band of TPLOs is fully occupied, resulting in a quantum spin Hall Mott insulator. A $U(1)$ spin rotation symmetry quantizes and protects the (many-body) spin Chern number in the AFM Mott phase.

To show that the above phase diagram is explicitly captured via Mott states forming in the SCWOs, we decompose the many-body ground state and its excitations in the partial Wannier basis for a doubled unit cell.
Starting from a supercell Bloch Hamiltonian (Eq. \ref{eqn:supercell-bloch-hamiltonian}) for the BHZ model with a doubled diamond unit cell, identical to Fig. \ref{fig:wannier-heuristic}(b), we construct a supercell Wannier basis using the projector method. The two trial wavefunctions are delta functions in real-space, placed on the two tiling vector induced sublattice sites in the supercell. Their weight is solely on the red orbitals with on-site potential $-M$ [Fig. \ref{fig:BHZ}(a)]. These trial wavefunctions are projected into the pair of folded supercell valence bands and made orthonormal via Gram-Schmidt. 

By construction, the first resulting state is a smooth, non-vanishing function over the entire Brillouin zone. Its Fourier transform defines the SCWO, depicted in Fig. \ref{fig:BHZ}(f). The choice of the delta function trial wave function yields an SCWO that is very close to maximal localization upon further gauge choice optimization, and we therefore adopt it for the analysis below. The second (orthogonal) state however retains the topological obstruction and exhibits a phase vortex-like non-analyticity; its Fourier transform consequently defines the TPLO [\ref{fig:BHZ}(g)]. The pair of SCWO and TPLO are centered on the two maximally-symmetric Wyckoff positions in the supercell. The real-space decay of the total per-site amplitude of the Wannier functions is depicted in Fig. \ref{fig:BHZ}(h).

\begin{figure*}[hbt!]
    \centering
    \includegraphics[width=\linewidth]{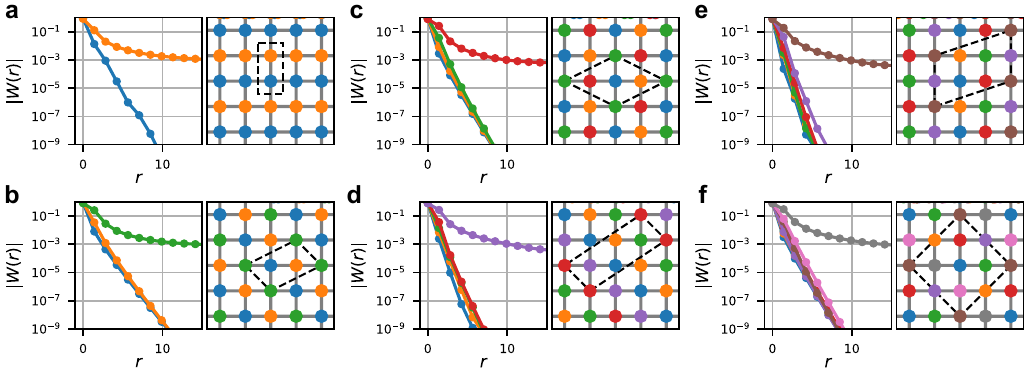}
    \caption{\captiontitle{Hierarchy of Partial Wannier Bases and Emergent Kondo Lattices.} \textbf{(a)}-\textbf{(f)} SCWOs and TPLOs for supercells on the square lattice containing $N=2,3,4,5,6,8$ unit cells, computed for the topological valence band of the BHZ model on a $40 \times 40$ mesh. Left panels depict the amplitude of SCWOs and TPLO as a function of distance along the $A_{1}$ direction with respect to the Wannier center. Right panels depict the supercell. Colors indicate the location of the Wannier centers.}
    \label{fig:supercellHierarchy}
\end{figure*}

To verify that the Mott insulating phases originate from the localization of single electron spins on the SCWOs, we now construct real-space SCWO fermion operators $\WD_{\R, \sigma} = \sum_{\R', \alpha, \t} W_{\alpha, \t}(\R - \R') \cD_{\R'+\t, \alpha, \sigma}$ where $W_{\alpha,\t}(\R)$ is the SCWO wave function [Fig. \ref{fig:BHZ}(f)], $\R$ indexes the supercells and $\t$ are tiling vectors. At $\nu = 1/2$, a single charge per supercell is heuristically expected to sit in the SCWO, forming a local magnetic moment that can order at low energies. To probe this, the SCWO fermion operators define a \textit{magnetic} (Ising) basis
\begin{align}
    \label{eqn:many-body-wannier-tiling}
    \ket{\Psi(\{\sigma\})} = \prod_{\R} \WD_{\R, \sigma(\R)} \ket{0}
\end{align}
for the emergent Kondo lattice. Here, $\ket{0}$ is the vacuum state and $\{\sigma\}$ are all $2^8$ possible spin-$z$ configurations for 8 electrons with spin $\uparrow$ or $\downarrow$ in the 8 SCWOs (at position $\R$) of the 16-site lattice that we consider.

Deep in the Mott phase, the magnetically-ordered ground states and spin excitations of the $1/2$-filled spin Chern band should be well-described (up to weak charge fluctuations) as a superposition of $\ket{\Psi(\{\sigma(\R)\})}$. To quantify this behavior, we consider the flat-band limit $\Gamma = 0$ with interactions $U'/U = 0.4$ and study the projection
\begin{align}
    f_n = \sqrt{ \bra{n} \left[ \sum_{\{\sigma\}} \ket{\Psi(\sigma(\R))} \bra{\Psi(\sigma(\R))} \right] \ket{n} }
\end{align}
of the $n < 1200$ low-energy many-body states $\ket{n}$ into this magnetic basis. This supercell Wannier subspace overlap $f_n$ approaches one if charge fluctuations are small and the partial Wannier basis captures the formation of local magnetic moments.

Fig. \ref{fig:BHZ}(i) shows $f_n$ for the lowest $n < 1200$ many-body eigenstates computed for the AFM Mott phase. In numerics, the anticipated checkerboard pattern of local moments comes with a two-fold degeneracy from spontaneous translation symmetry breaking, related via a simple lattice translation. Eigenstates in finite-size ED will always be Schr\"odinger cat superpositions of the translation symmetry broken partners. However, the trial wavefunction enforces an assumption about the charge ordering pattern from spontaneous symmetry breaking, i.e. the location of the SCWO in the supercell. To resolve this, a small bias $V_{b} \sim 0.05 U$ is added in ED to favor the symmetry-broken configuration that was enforced by the ansatz [Fig. \ref{fig:BHZ}(i), inset]. This bias is much smaller than the gap for charge excitations. We find that the first $256$ states exhibit $f_n \sim 95\%$ overlap with the Wannier subspace, corresponding precisely to the $2^8$ magnetic configurations for the SCWO local moments. The next $256$ states have vanishingly small projection into the SCWO Wannier subspace. Physically, these are the translation-symmetry-broken partners that are energetically separated by the bias potential. States beyond the $512$th eigenstate describe charge excitations with double occupancies, with small but non-zero Wannier subspace overlaps. In combination, we find that this Mott AFM phase is therefore well-described via a Mott-Hubbard model formed by the SCWOs, with TPLOs unoccupied below the charge gap.

Importantly, the choice of $1/2$ and $3/2$ filling is not unique. Instead, Kondo lattices with spontaneously broken translation symmetry can, in principle, emerge at near rational filling $p/q$, with a $q$-fold enlarged supercell. To illustrate this, Fig. \ref{fig:supercellHierarchy} depicts the localization properties of SCWOs and a single TPLO, for six choices of supercells containing $q=2,3,4,5,6,8$ unit cells, computed for the BHZ model with $q$ delta function trial states localized on the $q$ sites in the supercell. Remarkably, the $q-1$ SCWOs remain tightly localized for arbitrarily larger supercells, making them an ideal starting point to study a hierarchy of translation symmetry breaking Mott states and emergent Kondo lattices, as well as their stability upon doping.

\subsection{Twisted Bilayer MoTe\textsubscript{2}}

\begin{figure*}[hbt!]
    \centering
    \includegraphics[width = \linewidth]{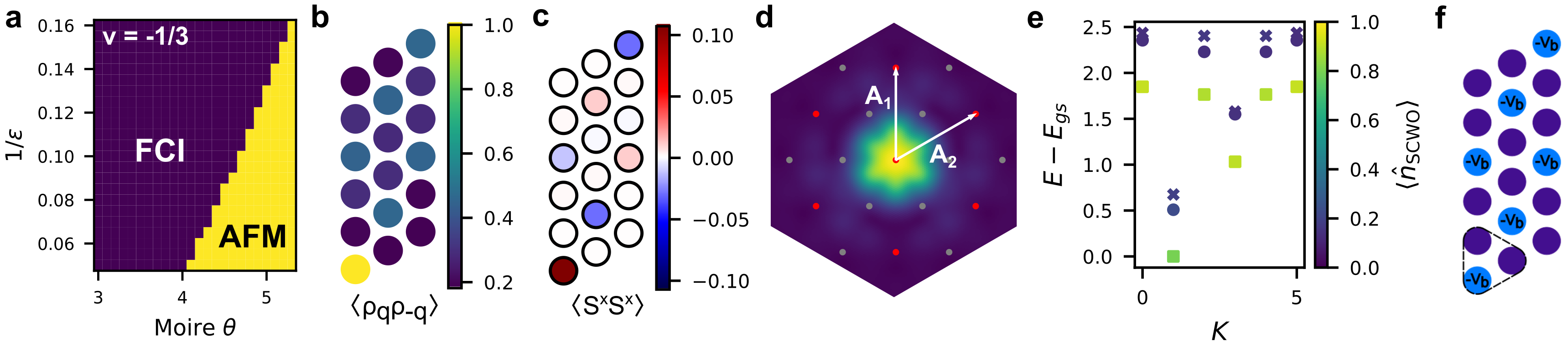}
    \caption{\captiontitle{Emergent Kondo lattices in twisted bilayer MoTe\textsubscript{2}.} \textbf{(a)} The phase diagram obtained at $\nu = -1/3$ as a function of the effective dielectric constant and moir\'e twist angle, showing an AFM Mott insulator on an emergent Kondo lattice, in competition with a $\nu=1/3$ Laughlin FCI \textbf{(b)} Representative density-density correlation function in the AFM region. \textbf{(c)} Representative $S^{x}S^{x}$ correlation function for the AFM phase, showing signatures of striped antiferromagnetic order. \textbf{(d)} The SCWO for the tripled $\sqrt{3} \times \sqrt{3}$ unit cell for the valence band. \textbf{(e)} The many-body spectrum obtained per total momentum $\K$ in the rBZ, by performing exact diagonalization on the interacting model in supercell Wannier basis. The color scale describes the number of electrons in the SCWO (d) per supercell. \textbf{(f)} A small bias field is added to single out one translation-symmetry-breaking pattern in computing (e).}
    \label{fig:tMoTe2}
\end{figure*}

Twisted bilayers of MoTe\textsubscript{2} are an ideal platform to observe this state of matter. At small twist angles, they host nearly-dispersionless and energetically-separated moir\'e valence bands with a non-zero spin Chern number which are composed of the two layers' valleys, locked to spin by virtue of the large Mo spin-orbit interactions~\cite{alidoust_observation_2014}. They also have been observed to stabilize ferromagnetic FCI states at $\nu = -2/3$ fractional filling~\cite{park_observation_2023, zeng_thermodynamic_2023, redekop_direct_2024, cai_signatures_2023}. However, competing and thus far unidentified correlated insulating phases without spin polarization have also been observed at other fillings, including $\nu = -1/3$~\cite{cai_signatures_2023}. 

To study the possibility for the formation of an antiferromagnetic-ordered emergent Kondo lattice, we study the interacting top-most spin Chern band of twisted MoTe\textsubscript{2} in ED and via construction of a supercell Wannier basis. We start from the single particle continuum model for a single valley locked to spin ~\cite{wu_topological_2019}
\begin{align}
    \mathcal{H} = \left[\begin{array}{cc} -\frac{\hbar^2 \left( -i\boldsymbol{\nabla} - \mathbf{K}_b \right)^2}{2m^*}  + V_b(\mathbf{r}) & V_T(\mathbf{r}) \\ V_T^*(\mathbf{r}) & -\frac{\hbar^2 \left( -i\boldsymbol{\nabla} - \mathbf{K}_t \right)^2}{2m^*} + V_t(\mathbf{r}) \end{array}\right] .
\end{align}
Here, $\mathbf{K}_b$ and $\mathbf{K}_t$ are $K$ points for the bottom and top layers and $m^* = 0.6 m_e$ is the effective mass. Electrons in the top and bottom layer experience moir\'e potentials $V_{t,b}(\mathbf{r}) = 2v \sum_{j=1,3,5} \cos(\mathbf{G}_j\cdot \mathbf{r} + \psi_{t,b})$, where $\mathbf{G}_j$ are the moir\'e reciprocal lattice vectors obtained via counter-clockwise rotation of $G_1 =(4\pi)/(\sqrt{3}a_M)[0,1]^\top$. Here, $a_M = a_0/\theta$ is the moir\'e length for twist angle $\theta$ where $a$ is the monolayer lattice constant, and $\psi_{b} = -\psi_{t} \equiv \psi$ is a model parameter. Interlayer tunneling is described by $V_T(\mathbf{r}) = w(1 + e^{-i\mathbf{G}_2 \mathbf{r}} + e^{-i \mathbf{G}_3 \mathbf{r}})$. The Hamiltonian for spin-$\downarrow$ is obtained via time reversal. We adopt parameters from Ref.~\cite{wang_fractional_2024} with $v = 20.8{\rm meV}$, $w=-23.8{\rm meV}$, and $\psi = 107.7^\circ$. 

We start from a screened Coulomb interaction
\begin{align}
    \hat{H}_{I} = \sum_{\substack{\k, \k', \q, \\ l, l', \sigma, \sigma'}} \frac{V(q)}{2A} \cD_{\k + \q, l, \sigma} \cD_{\k' - \q, l', \sigma'} \c_{\k', l', \sigma'} \c_{\k, l, \sigma}.
\end{align}
with $V(q) = e^{2} \tanh(|q|d)/2 \epsilon_{0} \epsilon |q|$ , where $l$, $l'$ are the layer indices, $e$ is the electron charge, $\epsilon$ is the relative dielectric constant, $d$ is the distance between the MoTe$_{2}$ sample and the metal gates, and $A$ is the area of the system. We choose a $6 \times 3$ torus, and again project the Coulomb interaction into the top-most time-reversed pair of Chern $C=1$ valence bands.

Fig. \ref{fig:tMoTe2}(a) depicts the phase diagram for twisted bilayer MoTe\textsubscript{2} at $\nu = 1/3$ hole filling, as a function of the twist angle and dielectric constant $\epsilon$. Small twist angles stabilize an FCI corresponding to a $\nu=1/3$ Laughlin state, which is identified via full spin polarization and a well-defined 3-fold quasi-degeneracy that is well-separated from excitations. Its phase boundary is determined via the second derivative of the ground state energy as a function of model parameters. At larger twist angles, however, we find a competing insulating state with $\sqrt{3} \times \sqrt{3}$ charge order [Fig. \ref{fig:tMoTe2}(b)] but no net spin polarization. The spin correlation functions for this phase are depicted in Fig. \ref{fig:tMoTe2}(c), showing signatures of striped antiferromagnetic order. The magnetic moments are localized on a $\sqrt{3} \times \sqrt{3}$ pattern. These observations are immediately consistent with the formation of an AFM Mott insulating state on an emergent Kondo lattice with a single SCWO in a $\sqrt{3} \times \sqrt{3}$ unit cell. We note that this region of the phase diagram was previous studied in Ref.~\cite{yu_fractional_2023}, who also point out that the FCI is destabilized by interband mixing with the second moir\'e valence bands. However, the spin-unpolarized AFM Mott state was not identified. By computational necessity, the choice of torus breaks rotational symmetry, which can artificially favor striped AFM over competing N\'eel states. 

We now demonstrate that the low-energy physics of this phase is well-described by local moments forming in a supercell Wannier basis for an emergent Kondo lattice. To construct the SCWOs, we build a supercell continuum model for tMoTe$_{2}$ using a $\sqrt{3} \times \sqrt{3}$ supercell, and compute the top three folded Bloch states $\ket{u_{n\K\sigma}(\mathbf{r})}$ per spin $\sigma$ that constitute the spin Chern valence bands. At $\nu = -1/3$ filling, a Mott-insulating state with $1/6$ hole per spin per moir\'e unit cell demands a single SCWO to hold the charge. We therefore choose a single-particle trial wave function $\ket{t(\mathbf{r})}$ that is a delta function localized in the AA stacking region and with weight solely in the top layer. We project the trial wave function onto the top three folded Bloch states $\ket{\Psi_{\K\sigma}(\mathbf{r})} = \hat{P}_{\K\sigma} \ket{t(\mathbf{r})} / \sqrt{ \bra{t(\mathbf{r})} \hat{P}_{\K\sigma} \ket{t(\mathbf{r})} }$ where $\hat{P}_{\K\sigma} = \sum_{n=1}^3 \ket{u_{n\K\sigma}(\mathbf{r})}\bra{u_{n\K\sigma}(\mathbf{r})}$, and obtain a single exponentially-localized SCWO after Fourier transform $\ket{\Psi_{\R\sigma}(\mathbf{r})} = \frac{1}{\sqrt{L}} \sum_\K e^{-i\K(\r-\R)} \ket{\Psi_{\K\sigma}(\mathbf{r})}$. The SCWO is shown in Fig. \ref{fig:tMoTe2}(d). The gauge choice can be slightly optimized to yield a minimum mean squared spread; however, the Wannier function from the above procedure is already well-localized in the AA regions of the twisted bilayer, and we adopt it henceforth.

We now compute the per-supercell electron occupancy in the SCWO for the ground state and low-energy excitations, by studying the expectation value
\begin{align}
    \hat{n}_{\text{SCWO}} = \frac{1}{L} \sum_{\K, \sigma} \hat{n}_{\K, \textrm{SCWO}, \sigma}
\end{align}
where $L$ is the number of supercells, and $\hat{n}_{\K,\textrm{SCWO},\sigma}$ is the number operator for the SCWO orbitals, in momentum space. To compute this efficiently, we complete the supercell single-particle basis by constructing two additional (arbitrary) projected trial states $\ket{\Phi_{2,\K\sigma}(\mathbf{r})}, \ket{\Phi_{3,\K\sigma}(\mathbf{r})}$, which are orthonormalized with respect to $\ket{\Psi_{\K\sigma}(\mathbf{r})}$ via the Gram-Schmidt procedure. The many-body supercell Hamiltonian in momentum space (after projection onto the top-most spin Chern bands) can now be rotated unitarily into this basis, and $\hat{n}_{\text{SCWO}}$ can be efficiently computed in momentum space. We note that for computational expediency, this analysis differs in detail from the many-body SCWO subspace projection used for the BHZ model. In contrast to the computation of $f_n$, it permits leveraging the (supercell) translation symmetry in ED, to make simulations on $6 \times 3$ clusters feasible. The other two trial wavefunctions are largely irrelevant to the analysis outside of ensuring unitarity of the basis rotation.

Fig. \ref{fig:tMoTe2}(e) shows the number of electrons per SCWO per supercell, computed for the lowest-energy many-body states for twist angle $\theta = 5^\circ$ and $\epsilon = 16.7$, deep in the AFM phase, and indexed by the total supercell momentum $\K$. $\langle \hat{n}_{\text{SCWO}} \rangle \to 1$ indicates that all electrons are localized in the SCWO and form a local moment on the emergent Kondo lattice. Analogous to the BHZ model, the Schr\"odinger cat superposition of translation-symmetry-broken partners is a finite-size effect. We again introduce a small artificial bias term to favor the gauge's chosen translation-symmetry-broken configuration [Fig. \ref{fig:tMoTe2}(f)]. This bias is much smaller than both the charge gap and energy scale for spin excitations. In the rBZ, the almost-threefold-degenerate many-body eigenstates correspond to the three translation-symmetry-broken partners. They occupy the same total supercell momentum sectors, since the tiling vectors are sublattice indices on the same supercell site in momentum space. With a small bias, the spectrum in each momentum sector therefore splits into triplets, related via elementary moir\'e lattice translations. The lowest energy eigenstate per triplet in each momentum sector has almost unit occupancy of the SCWO, confirming the formation of a Mott-like state on a $\sqrt{3} \times \sqrt{3}$ emergent Kondo lattice.

\section{Conclusion \& Outlook}

In this work, we have shown that strong interactions in fractionally-filled topological bands can lead to the formation of \textit{emergent} Kondo lattices hosting a new class of Mott-insulating states. We have demonstrated that these phases can emerge in twisted bilayers of MoTe\textsubscript{2} as well as the fractionally-filled BHZ models. We found that they are naturally described via conventional Mott-Hubbard interactions in a partial supercell Wannier basis for Chern bands, with a gauge choice that relaxes translation symmetry. Such partial Wannier bases are rooted in the observation that an $N$-band manifold with net non-zero Chern number permits the construction of up to $N-1$ exponentially-localized Wannier orbitals, with solely a single residual state retaining the original net Chern number. In time-reversal-symmetric spin Chern bands, this leads to the formation of Kondo lattices with magnetic moments localized on tightly-localized Wannier orbitals and coupled to an empty or dilutely-filled itinerant topological band from residual extended orbitals with power-law localization that carry the topological invariant.

Our results both establish a new theoretical tool and predict a class of topological Mott states in Chern bands that can be readily observed in twisted transition-metal dichalcogenide heterostructures. The predicted states of matter constitute a competing phase for fractional Chern insulator and realize a new scenario for time-reversal-symmetric interacting topological bands in solids, beyond conventional Landau level paradigms for the fractional quantum Hall effect.

We note that the supercell Wannier basis approach for isolated Chern bands is different from established approaches, as only the active degrees of freedom in the isolated Chern bands are used. This contrasts with the topological heavy fermion model for twisted bilayer graphene~\cite{PhysRevLett.129.047601}, which augments the active electronic states of the fragile topological flat bands with higher-energy inert single-particle states to avoid the topological obstruction. Extensions to multiple active bands are straightforward and the presented partial supercell Wannier basis construction can similarly provide an ideal basis to understand translation-symmetry-breaking incompressible and topological Mott states. Furthermore, it will be interesting to investigate the magnetic interactions between SCWOs of the emergent Kondo lattice at strong coupling, for possible orders. Effective exchange interactions can arise via superexchange between neighboring SCWOs, however, RKKY interactions mediated via the extended TPLOs can lead to richer magnetic phases from longer-ranged interactions.

At fractional commensurate fillings in an interacting system, the incomplete SCWO basis together with the TPLO state become a very natural choice of describing broader classes of correlated insulating states that spontaneously break translation symmetry. For instance, in addition to the topologically trivial and QSH antiferromagnetic systems discussed in this work, we expect that this basis is a very natural choice for describing a whole host of generalized Wigner crystals in topological bands~\cite{regan_mott_2020, li_imaging_2021, jin_stripe_2021, han_correlated_2023}. Similarly, it will be interesting to apply the present analysis to other kinds of interacting topological bands, with strong or fragile symmetry-protected topological obstructions. For instance, the partial supercell Wannier construction forms a natural starting point for understanding possible nematic or charge-ordered phases in the flat bands of twisted bilayer and multilayer graphene.

Studying quantum anomalous Hall crystals~\cite{sheng_quantum_2024, perea-causin_quantum_2024, zhou_new_2024} from the perspective of partial supercell Wannier bases raises intriguing questions about the structure of the single-particle many-body spectral function, beyond weak-coupling Hartree-Fock descriptions. To obtain a fully spin-polarized state with a non-zero charge Chern number for $\nu = -1/2$ hole filling, the TPLOs must be occupied by holes with a single spin polarization, while the SCWOs need to remain empty of holes, in contrast to the Mott states discussed here. This is possible in principle if the interactions are very well-screened and sufficiently weak such that kinetic energy gains from hoppings from the TPLOs become prominent. Studying the competition between a QAH crystalline state and the strong-coupling emergent Kondo lattice phases predicted here will be an interesting topic for future work.

Topologically, even richer possibilities exist for emergent Kondo lattices with AFM order in fractionally-filled topological bands in three dimensions. In a 3D topological insulator with a fractionally-filled $\mathbb{Z}_2$ topological band, an emergent Kondo lattice with AFM order can become a magnetic topological Mott insulator with an axion electrodynamical response~\cite{PhysRevB.78.195424,PhysRevLett.102.146805,rundongNaturePhys2010,magneticTopoReview}. The partial supercell Wannier construction and its localization properties in three dimensions can provide a guiding principle for assessing candidate materials.

Finally, it will be interesting to understand the stability of emergent Kondo lattices in a topological band upon doping. As any rational filling $\nu = p/q$ in principle admits a Kondo lattice description with a $q$-fold enlarged supercell, there is an intrinsic competition between doping electrons into the itinerant band of TPLOs and transitioning to a different supercell of local moments and empty or dilutely-filled TPLOs. Studying the resulting hierarchy of stable filling fractions will be intimately tied to the localization properties of the partial Wannier basis and constitute an interesting direction for future research.

\acknowledgements{
    We are grateful for helpful comments and discussions with Charles Kane, Eugene Mele, Spenser Talkington, and Jixun K. Ding. We acknowledge support from the Alfred P. Sloan Foundation through a Sloan Research Fellowship and a startup grant from the University of Pennsylvania.
}

%\bibliographystyle{apsrmp} % We choose the "plain" reference style
%\bibliography{refs} % Entries are in the refs.bib file

\end{document}